\documentclass[twocolumn]{ase}

\usepackage{cite}
\usepackage{amsmath}   
\usepackage{mathtools}  
\usepackage{amsfonts}  
\usepackage{url}     
\usepackage{color}     
\usepackage{hyperref} 

\usepackage{graphicx}
\usepackage{booktabs, multicol, multirow}
\usepackage{listings}
\usepackage[usenames,dvipsnames]{xcolor}
\usepackage{tikz}
\usepackage{tabularx}
\usepackage{booktabs}
\usepackage{threeparttable}
\usepackage[T1]{fontenc}
\usepackage{tgpagella}
\usepackage{dtklogos}
\usepackage[ruled,vlined]{algorithm2e} %

\DeclareMathOperator*{\argmax}{\arg\!\max}

\begin{document}
\sloppy
\hfuzz=1.5pt

\twocolumn
[\begin{center}
{\huge Automatic Labeling for Entity Extraction in Cyber Security} 



\vspace{.15in}
{Robert A. Bridges$^1$, Corinne L. Jones$^2$, Michael D. Iannacone$^3$, Kelly M. Testa$^4$, John R. Goodall$^5$}\\
$^1$bridgesra@ornl.gov, $^2$corinne.jones180@gmail.com, $^3$iannaconemd@ornl.gov, $^4$testakm@ornl.gov, $^5$jgoodall@ornl.gov \\
Cyber \& Information Security Research Group\\
Oak Ridge National Laboratory\\
Oak Ridge, TN 37830

\end{center}]

\begin{abstract}
Timely analysis of  cyber-security information  necessitates automated information extraction from unstructured text.  
While state-of-the-art extraction methods produce extremely accurate results, they require ample training data, which is generally unavailable for specialized applications, such as detecting security related entities; moreover, manual annotation of corpora is very costly and often not a viable solution.
In response, we develop a very precise method to automatically label text from several data sources by leveraging related, domain-specific, structured data and provide public access to a corpus annotated with cyber-security entities.
Next, we implement a Maximum Entropy Model trained with the average perceptron on a portion of our  corpus ($\sim$750,000 words) and achieve near perfect precision, recall, and accuracy, with training times under 17 seconds. 
\end{abstract}

\section{Introduction}
\label{intro}
Online security databases, such as the National Vulnerability Database (NVD), the Open Source Vulnerability Database (OSVBD), and  Exploit DB are important sources of security information, in large part because their well defined structure facilitates quick acquisition of information and allows integration with various automated systems.\footnote{\url{http://nvd.nist.gov/}, \url{http://www.osvdb.org/}, \url{http://www.exploit-db.com/}}  
On the other hand, newly discovered information often appears first in unstructured text sources such as blogs, mailing lists, and news sites.
Hence, in many cases there is a time delay, sometimes months, between public disclosure of  information and appropriate classification into structured sources (as noted in \cite{mcneil2013pace}).  
Additionally, many of the structured sources include a text description that provides important details (e.g., Exploit DB).   
Timely use of this information, both by security tools and by the analysts themselves, necessitates automated information extraction from these unstructured text sources.

For identifying more general entity types, many ``off-the-shelf'' software packages give impressive results using proven supervised methods trained on enormous corpora of labeled text.   
Because the training data is only annotated with names, geo-political entities, dates, etc., these general entity recognition tools are inadequate when expected to extract the relatively foreign entities that occur in domain-specific documents, simply because they are not trained to handle such jargon. 
Exemplified by our need for entity extraction in the cyber-security domain, there are many domain-specific applications for which entity extraction will be very beneficial. 
As evidenced by the near perfect results of sequential labeling techniques, for example \cite{manning2011part}, the machine learning is thoroughly developed.  
Rather, what is lacking is labeled training data tailored to domain specific needs.  
Moreover, manual annotation of a sufficiently large amount of text is generally too costly to be a viable solution.   

This paper describes an automated process for creating an annotated corpus from text associated with structured data that can produce large quantities of labeled text relatively quickly (compared to manual annotation) by writing a script which labels text with related structured sources.  
More specifically, the wealth of structured data available in the cyber-security domain is leveraged to automatically label associated text descriptions and made publicly available online.\footnote{\url{https://github.com/stucco/auto-labeled-corpus}}
While labeling these descriptions may be useful in itself, the intended purpose of this corpus is to serve as training data for a supervised learning algorithm that accurately labels other text documents in this domain, such as blogs, news articles, and tweets.  

Next, we use a portion of the data to train a history-based Maximum Entropy Model with the averaged perceptron and greedy decoding, and exhibit precision, recall, and accuracy that are consistently above 97\%; moreover, the algorithm runs extremely efficiently, training on over $750,000$ labeled words in under 17 seconds.  
In Section \ref{sec:results}, we compare our work to a previous similar attempt (\cite{joshiextracting} ) at supervised entity extraction within the cyber-security domain, which produced scores under 80\% when trained on a hand-labeled corpus.  While this is not a direct comparison, the increase in performance is evidently in part due to the vast increase in training data as facilitated by our automated labeling process.    


\section{Background}
\subsection{Entity Extraction in Cyber-Security Overview}
Our overall goal of automatically labeling cyber security entities is similar to a few previous efforts.
In order to instantiate a security ontology, More et al. \cite{more2012knowledge} attempt to annotate concepts in the Common Vulnerability Enumeration (CVE)\footnote{\url{http://cve.mitre.org/}} descriptions and blogs with OpenCalais, an ``out-of-the-box'' entity extractor \cite{reuters2009opencalais}.  
Mulwad et al. \cite{mulwad2011extracting} expand this idea by first crawling the web and training a decision classifier to identify security relevant text.  Then using OpenCalais along with the Wikipedia taxonomy, they identify and classify vulnerability entities.

\begin{figure}
\centering
\framebox{
\includegraphics[width=3.1in]{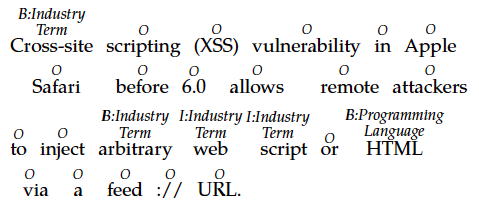}
}
\caption{NVD text description of CVE-2012-0678 with labels from OpenCalais.}     
\label{calais_eg}
\end{figure}

While the two sources above rely on standard entity recognition software, such tools are not trained to identify domain specific concepts, and they unsurprisingly give poor results when applied to more technical documents (as shown in Figure \ref{calais_eg}).  
This is due to the general nature of their training corpus; for example, the Stanford Named Entity Recognizer\footnote{\url{http://nlp.stanford.edu/software/CRF-NER.shtml}} is trained on the CoNLL, MUC-6, MUC-7 and ACE named entity corpora, consisting of news documents annotated mainly with the names of people, places, and organizations \cite{Finkel:2005:INI:1219840.1219885, tjong2003introduction}.
Similar findings are noted in Joshi et al.'s recent work \cite{joshiextracting}, where OpenCalais, The Stanford Named Entity Recognizer, and the NERD framework \cite{rizzo2012nerd} were all generally unable to identify cyber security domain entities.
Because these tools do not use any domain specific training data,  domain entities are either unrecognized or are labeled with descriptions that are too general to be of use (e.g., ``Industry Term'').
The Joshi et al. paper later supplies the Stanford Named Entity Recognizer framework with domain specific, hand labeled  training data, and it is then able to produce better results for most of their domain specific entity types. 
%

More specifically, the Joshi et al. \cite{joshiextracting} work also addresses the problem of entity extraction for cyber-security with a similar solution, namely, by training a supervised learning algorithm to identify desired entities.  
Unlike our approach, which introduces an automated way to generate an arbitrarily large training corpus,  
their approach, involves painstakingly hand-annotating a small corpus that is then fed into the Stanford Named Entity Recognizer's ``off-the-shelf'' template for training a conditional random field entity extractor \cite{Finkel:2005:INI:1219840.1219885}. 
 In all, they label a training corpus of 350 short text descriptions, mostly from CVE records, with categories surprisingly similarly to ours.
While their work has identified the same cyber-security problem, they do not furnish a  data set labeled for this domain, nor do they address the more general problem of how to automate the labeling process when no training data exists.  See Section~\ref{sec:results} for detailed comparisons of the results, and \cite{Lal:2013} for more specifics on the entity extraction implementation as used in the Joshi paper.  
 
Given this general lack of domain specific training data, there has been some work considering semi-supervised methods instead of supervised methods because they are designed to do the best possible with very little training data.
Although a thorough discussion of semi-supervised methods for entity extraction is outside the scope of the current paper, such techniques have yielded worthwhile results; for example see \cite{brin1999extracting, jones2005learning, carlson2009nell, carlson2010coupled, carlson2010active}, and \cite{riloff2012event}.  
To our knowledge only one such effort focuses on cyber-security; 
recent work by McNeil et al. \cite{mcneil2013pace}  develops a novel bootstrapping algorithm and describes a prototypical implementation to extract information about exploits and vulnerabilities. 
 Using known databases for seeding bootstrapping methods is also not uncommon; for example, see \cite{Geng:2004:AAE:1018432.1021517}.  
 
\subsection{Automatic Labeling Overview} 
Previous work has incorporated variations of auto-labeling in several different contexts where NLP is needed and no training data exists. 
``Distant labeling'' generally refers to the process of producing a gazetteer (comprehensive list of instances) for each database field and performing a dictionary look-up to label text that is not directly associated with a given database record. 
While gazetteers give poor results in an unconstrained setting \cite{bird2009natural}, accurate results can be achieved when the text has little variation.       
An example is Seymore et al. \cite{Seymore99learninghidden} who use a database of \BibTeX \phantom{ } entries and some regular expressions to produce training data for a Hidden Markov Model (HMM) by labeling headers of academic papers.

In general, more accurate labels are possible if there is a direct relationship between a given database record and the text entry to be labeled, 
such as if a text description occurs as a field of a database, or a separate text document is referenced for each record, as is the case for our setting.   
Here we describe known instances of using an automated process for creating labeled training data.  
Craven and Kumlien \cite{Craven:1999:CBK:645634.663209} train a naive Bayes classifier to identify sentences containing a desired pair of entities via ``weak labeling''.  
Specifically, given a database record that includes a pair of entity names along with a reference to an academic publication, sentences occurring in the article's abstract are automatically labeled positively if that entity pair occurs in them.  
This is shown to yield better precision and recall scores than using a smaller hand-annotated training corpus and obviates the tedious manual labor. 

More recently, Bellare and McCallum \cite{bellare2007learning} also use a \BibTeX \phantom{ } database to label corresponding citations and then train a classifier to segment a given citation into authors, title, date, etc. 
Because their goal is to create a text segmentation tool, they rely on the implicit assumption that every token will receive a label from the given database field names.  
As our goal is to identify and classify specific entities in text, no such assumption can be leveraged.  


While a few instances of automated labeling have occurred in the literature, to our knowledge no previous work has addressed the accuracy of the automatically prescribed labels.  
Rather, an increase in accuracy of the supervised algorithm is usually attributed to the increase in training data, which is facilitated by the automated process.  
We note that the precision and recall of an algorithm's output is determined by comparison against the training data, which may or may not have correct labels.   
In order to address the quality of our auto-labeling, we have randomly sampled sentences for manual inspection (see Auto-Labeling Results Subsection \ref{autotag}.\ref{auto_results}).




\begin{figure}
\centering
\framebox{
\includegraphics[width=3.1in]{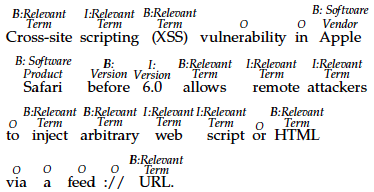}
}
\caption{NVD text description of CVE-2012-0678 with automatically generated labels.}     
\label{auto_tag_eg}
\end{figure}

\section{Automatic Labeling}
\label{autotag}
\subsection{Data Sources}
\label{data}

To build a corpus with security-relevant labels, we seek text that has a close tie to a database record and use its field names to label matching entries in the text.  
When a vulnerability is initially discovered, the Common Vulnerability Enumeration (CVE) is usually the first structured source to ingest the new information and it provides, most importantly, a unique identification number (CVE-ID), as well as a few sentence overview.  
Shortly afterward, the National Vulnerability Database (NVD) incorporates the CVE record and adds additional information such a classification of the vulnerability using a subset of the Common Weakness Enumeration (CWE) taxonomy\footnote{\url{http://cwe.mitre.org/}},  a collection of links to external references,  and other fields.
Hence, the NVD provides both informative database records and many structured fields to facilitate auto-labeling.  
All NVD descriptions from January 2010 through March 2013 have been auto-labeled and comprise the lion's share of our corpus.  

While our main source for creating an auto-labeled corpus is the NVD text description fields, the universal acceptance of the CVE-ID allows text from other sources to be unambiguously linked to a specific vulnerability record in the database.  
The Microsoft Security Bulletin provides patch and mitigation information and gives a wealth of pertinent text related to a specific vulnerability identified by the CVE-ID.\footnote{\url{http://technet.microsoft.com/en-us/security/bulletin}}  
Specific text fields include an ``executive summary'' as well as ``revision'', ``general'', ``impact'', ``target set'', ``mitigation'', ``work around'', and ``vendor fix'' descriptions; moreover, while not all text fields are populated for a given record, many times a single text field will have multiple descriptions.  
Every description for the previous year's MS-Bulletin entries was added to our corpus. 

Lastly, the Metasploit Framework contains a database of available exploits that includes a text description, several categorizations and properties, and a reference to the associated vulnerability, usually the CVE-ID.\footnote{\url{http://www.metasploit.com/}}
By linking these text sources to the NVD via CVE-IDs we are able to leverage the structured data for very precise labeling of the unstructured data.  

Overall, a corpus of over 850,000 tokens with automatic annotations are available online at \url{https://github.com/stucco/auto-labeled-corpus}. 

\subsection{Auto-Labeling Details}
\label{auto_details}
Given a database record and a block of associated text, our algorithm assigns labels to the entities in the text as follows:
\begin{itemize}
\item {\bf Database Matching.} 
Any string in the text that exactly matches an entry of the database record is labeled with a generalization of the name of the database field.  For example, the label ``software product'' is assigned to a string in the text description if it also occurs in the related database record field  ``os'' or ``application''.  Similarly, instances of ``version'', ''update'', and ``edition'' occurring in the associated text are labeled ''software version''.  
\item {\bf Heuristic Rules.} 
A variety of heuristic rules are used for identifying entities in text that are not direct matches of database fields.
For example, the database lists every version number affected by a vulnerability, but such a list is almost never written in text; rather, short phrases such as ``before 2.5'', ``1.1.4 through 2.3.0'', and ``2.2.x'' usually appear after a software application name; consequently, a few regular expressions combined with rules identifying both labels and features of previous words give precise identification of version entities.   
Similarly, source code file names, functions, parameters, and methods, although not in the database, are often referenced in text.  
As file names end in a file extension (e.g., ``.dll'') and the standards of camel- and snake-case (e.g., camelCaseExample,  snake\_case\_example) are universal, such entities are easily distinguishable by their features.
\item {\bf Relevant Terms Gazetteer.}
In order to extract short phrases that give pertinent information about a vulnerability, a gazetteer of relevant terms is created, and phrases in the text matching the gazetteer are labeled ``relevant term''.  
As mentioned above, each record in the NVD includes one (of twenty) CWE classifications, which gives the vulnerability type (e.g., SQL injection, cross-site scripting, buffer errors).  
As the goal of CWE is to provide a common language for discussing vulnerabilities, many phrases indicative of the vulnerability's characteristics occur regularly.  
To construct the gazetteer of relevant terms, the NVD is sorted by CWE type, and statistical analysis of the text descriptions for a given CWE classification is used to find the most prevalent unigrams, bigrams, and trigrams.   
Commonly occurring but uninformative phrases (e.g., ``in the'', ``is of the'') are discarded manually.
We note that Python's Natural Language Toolkit (NLTK) facilitated tokenization and computation of frequency distributions of $n$-grams~\cite{bird2009natural}.  
Examples of relevant terms include ``remote attackers'', ``buffer overflow'', ``execute arbitrary code'', ``XSS'', and ``authentication issues''.
\end{itemize}

All together, the following is the comprehensive list of labels used: ``software vendor'', ``software product'', ``software version'', ``software language'', ``vulnerability name'' (these are CVE-IDs), ``software symbol'' (these are files, functions or methods, or parameters), and ``vulnerability relevant term''.


Because many multi-word names are commonplace, standard IOB-tagging is used; specifically, the first word of an identified entity name is labeled with a ``B'' (for ``beginning'') followed by the entity type, and any word in an entity name besides the first is tagged with an ``I'' (for ``inside'') followed by the entity name.  
Unidentified words are labeled as ``O''.  
An example of an automatically labeled NVD description is given in Figure \ref{auto_tag_eg}.  

\subsection{Auto-Labeling Results}
\label{auto_results}
As the overall goal is to produce a machine learning algorithm that will identify entities in a much broader class of documents, thereby aiding security analysts, the accuracy of the algorithm, and therefore the training data, is very important.  
While both high precision and recall are ideal, precision is more important for our purposes as reliable information is mandatory.  
More specifically, in a high recall but low precision setting, nearly all desired entities would be returned along with many incorrectly labeled ones; hence, the quality of the data returned to the user would suffer.
On the other hand, if all information extracted from text sources is correct, anything returned is an immediate value-add. 
In general, this is guaranteed by high precision in the auto-labeling process, which we ensure by constructing precise heuristics and using a specific database record to label closely related text.  

\begin{center}
\begin{table}[h]
\begin{tabular}{|l|l|l|l|}
\hline
            & \textbf{Precision} & \textbf{Recall} & \textbf{F1}     \\ \hline
\textbf{NVD}         & 99\%      & 77.8\% & .875 \\ \hline
\textbf{MS-Bulletin} & 99.4\%    & 75.3\% & .778   \\ \hline
\textbf{Metasploit } & 95.3\%    & 54.3\% & .691  \\ \hline
\end{tabular}
\caption{Precision, Recall, and F1 Scores for the automatically labeled corpus are calculated by hand labeling a random sample.}     
\label{auto_tag_results_table}
\end{table}
\end{center}
In order to test the accuracy of the auto-labeling, about 30 randomly sampled text descriptions from each source were manually labeled.
Because the label ``relevant term'' is applied by a direct dictionary look up against a list of terms we created, we know each and every exact match in the text is labeled; hence, they are not included in the accuracy scores to prevent artificial score inflation.
In other words, the Precision, Recall, and F1 Score results of Table \ref{auto_tag_results_table} are with respect to only those labels matching an entry of a database field or from a hand-crafted heuristic.    


To our knowledge, similar work has assumed correct automatically generated labels and ignored investigating the accuracy of the labels.  
In total over 850,000 tokens have been labeled relatively quickly (with respect to manual annotation) and with high accuracy, and increasing the corpus size as necessary is both expedient and easy.  
We hope the proposed method can facilitate labeling data in many other domains.  
\section{Entity Extraction Via Sequential Labeling}
As is common in the literature, our approach to supervised entity extraction is treating the task as a sequential labeling problem, similar to previous work on part-of-speech tagging, noun phrase chunking, and parsing.  
This section gives an overview of machine learning techniques for such a task and reviews the mathematical foundation for Maximum Entropy (or Log-Linear) Models in preparation for our implementation, described in the Section \ref{implementation}.  

\subsection{Sequential Tagging Models}
Used widely in sequential tagging problems, Hidden Markov Models (HMMs) are generative models that estimate the joint probability of a given sentence and corresponding tag sequence by first estimating an emission parameter, that is, the probability of a word given its label, and secondly, by estimating a prior distribution on the set of labels using a Markov Process \cite{Collins:2002:DTM:1118693.1118694}.    

While HMMs are computationally efficient, the subclass of discriminative models known as Maximum Entropy Models (MEMs) are perhaps a more popular choice for sequential tagging problems as they generally  outperform Hidden Markov Models by virtue of their accommodation of a much larger set of features; for example, see \cite{McCallum:2000:MEM:645529.658277, Ratnaparki96}. 
Two varieties of MEMs are common in the literature, namely, those using ``history-based'' features (whose features depend on the current word as well as previous word(s) and label(s)) and those using ``global features'' (whose features depend on both the words and labels before and after a given word).  
More commonly referred to as Conditional Random Fields (CRFs), global models treat each sentence as an object to be labeled with a corresponding set of word tags (rather than labeling individual words sequentially) and have achieved better performance than history-based MEMs, but at the price of greater computational expense \cite{Lafferty:2001:CRF:645530.655813}. 
More specifically, with $k$ possible word labels and a sentence of length $n$, the search space for sentence tags is of order $n^k$. 
Because of the dependence only in the reverse direction, history-based MEMs admit use of the Viterbi algorithm for finding the most probable tag sequence efficiently (with order $nk^m$ for features depending on the previous $m$ labels); 
furthermore, one has the option of a greedy algorithm, which inductively chooses the highest probability tag for each word and ignores the overall probability of the sequence.  
As no such options exist for decoding with CRFs, efforts include incorporating an algorithm for narrowing the search space or using probabilistic means for finding the best tag sequence \cite{Collins:2002:DTM:1118693.1118694, Finkel:2005:INI:1219840.1219885}.  
Because of the observed performance of the history-based MEM with a greedy tagging algorithm in our setting (see Subsection~\ref{sec:results}), use of more computationally expensive algorithms, such as CRFs or even Viterbi decoding, was unwarranted.  
  
\subsection{Mathematical Overview} 
A brief mathematical overview of a history-based MEM is followed by  the implementation details used in our experiment.   

Derived by maximizing Shannon's Entropy in the presence of constraint equations, MEMs  provide a principled mathematical foundation that ensures only the observed features design the probability model. 
For a given sentence $w=(w_1,\dots, w_n)$ and corresponding tag sequence $t=(t_1,\dots,t_n)$, the conditional probability of $t$ given $w$ is estimated as 
\begin{equation}
\label{conditional_prob_w}
p(t|w)\equiv \prod_{i=1}^n p(t_i|t_{i-2}, t_{i-1}, w_{i-2}, w_{i-1}, w_i)
\end{equation}
with $t_0, t_{-1}, w_{0}, w_{-1}$  defined to be distinguished start symbols.  
Hence the probability of tag $t_j$ being assigned to word $w_j$  is conditioned on the previous two tags (in our implementation), as well as the current word and previous two words.  
For notational ease we let $\bar{t}_i=(t_{i-2},t_{i-i},t_i)$, and similarly for $\bar{w}$.  
As prescribed by the MEM, 
\begin{equation}
\label{conditional_prob}
p(t_i|t_{i-2}, t_{i-1}, w_{i-2}, w_{i-1}, w_i)\equiv \frac{e^{f(\bar{t}_i, \bar{w}_i)\cdot v}}{z(\bar{t}_i, \bar{w}_i)}
\end{equation}
where $f=(f_1,\dots,f_m)$ denotes a feature vector, \\
$v=(v_1,\dots,v_m)$ the parameter vector (or feature weights) to be learned from the training data, and 
$$z(\bar{t}_i, \bar{w}_i)\equiv \sum_{\hat{t}} \exp[{f(t_{i-2},t_{i-1},\hat{t}, \bar{w}_i)\cdot v}],$$
i.e., $z$ is the appropriate constant to ensure the sum of Equation~\ref{conditional_prob} over the sample space is one.   
An example of a feature  (i.e., a component of the feature vector) is
\begin{equation}
\label{feature_eg}
f_1(\bar{t}_i, \bar{w}_i)=\left\{ \begin{array}{cc}
1 & \mbox{ if } \begin{array}{c} t_i=\mbox{ B: SW Vendor}, \\ w_{i-1}=\mbox{ ``the''}\end{array} \\
0 & \mbox{ else. }
\end{array}\right.
\end{equation}
  
After fixing a set of features, one must decide on the ``best'' parameter vector $v$ to use and  many techniques for fitting the model to the training data (i.e., learning $v$)  exist \cite{elkan2008log}. 
Perhaps the most principled approach for fitting the model is maximum likelihood estimation (MLE), which assumes each (sentence, tag sequence)-pair is independent and uses a prior on $v$ (or regularization parameter) to prevent over-fitting. 
Specifically,  the argument maximum of
$$v\mapsto p(v|\{(w,t)\}) \propto \prod_{(w,t)}p((w,t)|v)p(v)$$ is generally found by  maximizing the log-likelihood, usually by a numerical algorithm such as L-BFGS, or  OWL-QN \cite{Andrew:2007:STL:1273496.1273501}.  
We note that the function in question is concave, and has a unique maximum.  

Initially introduced in \cite{rosenblatt1958perceptron}, the perceptron algorithm and its modern variants are a class of online methods for fitting parameters that have produced competitive results in accuracy and are often more efficient than MLE techniques\cite{Collins:2002:DTM:1118693.1118694, freund1999large}.  
After initializing the parameter vector $v$ (usually setting $v=0$), perceptron algorithms cycle through the training set a fixed number of times.  
At each training example the algorithm predicts the ``best'' label with the current parameter $v$ and compares it to the ground-truth value. 
 In the case of a mis-assigned label, the parameter $v$ is updated so that the probability of the correct label increases.   
As perceptron algorithms depend on decoding at each step, their computational expense can vary, but in the case of greedy or Viterbi decoding, they are relatively fast.

\section{Entity Extraction Implementation}
\label{implementation}
While the auto-labeled corpus may be useful in its own right, the overall goal is to train a classifier that can apply domain-appropriate labels to a wider class of documents including news articles, security blogs, and tweets.  
Our choices for such implementation follow Mathew Honnibal's persuasive results and documentation of greedy tagging using the averaged perceptron for part-of-speech tagging,\footnote{http://honnibal.wordpress.com/2013/09/11/a-good-part-of -speechpos-tagger-in-about-200-lines-of-python/} where he shows impressive results with respect to a balance of accuracy, speed, and simplicity.  
To our knowledge no publication of the results exists. 
Here we give a brief synopsis of possible tagging algorithms, and describe our implementation of a history-based Maximum Entropy Model trained with the averaged perceptron. 
Finally, we present performance results from a simple greedy model for tagging.
  
\subsection{Averaged Perceptron}  
We chose to use a modern perceptrion variant, namely,  the averaged perceptron, which has exhibited exceptional results in many natural language processing tasks  \cite{Collins:2002:DTM:1118693.1118694,  collins2004incremental,freund1999large, zhang2007chinese}.    
The averaged perceptron algorithm is presented in detail in Algorithm \ref{ave_per}, and explained below. 
 \begin{algorithm}
\SetEndCharOfAlgoLine{} 
\KwIn{$\{(w,t)\} = $ training set \\
$N_{iter} = $ number of iterations\\}
\KwOut{$v_{ave} = $ trained parameter vector}
Initialize $iter = 1$ \;
Initialize $i = 0$ \;
Initialize $v = (0,\dots, 0)$\;
Initialize $v_{t-stamp} = (0, \dots, 0)$ \;
Initialize $v_{tot} = (0, \dots, 0)$ \;
\While{$iter \leq N_{iter}$}{
  \For{$(w,t)$ in training set}{
	Set $y=\argmax_{\hat{t}} p(\hat{t}|w,v)$ \; 
    \uIf{$y!=t$ }{
      $v_{tot} += [(i, \dots, i)-v_{t-stamp}]*v$ \;
      $v += f(w,t) - f(w,y)$ \;
      \For{ $j = 1 \dots \mbox{ length}(v)$ such that $f(w,y)[j] != 0$}{
    		Set $v_{t-stamp}[j] = i$\;
    		}
    	 $i += 1 $ \;%
    }
    \Else{
      $i += 1$ \;
    }
  }
  $iter += 1$ \;
}
$v_{tot} += [(i, \dots, i)-v_{t-stamp}]*v$\;
Set $v_{ave} = v_{tot}/i$\;
\Return{$v_{ave}$}\;
\caption{{ \sc Averaged Perceptron}}
\label{ave_per}
\end{algorithm}

The averaged perceptron uses the same online algorithm to tweak the parameter vector as it iterates through the training set, although this updated vector from the ``vanilla'' perceptron training is not returned.  
Instead, we now keep track of how many successful labels are predicted by each intermediate parameter vector and  return the weighted average of the vectors observed in training.  
Rather than storing every intermediate vector along with a tally of each vector's success, the implementation below keeps two auxiliary vectors, a time-stamp ($v_{t-stamp}$), which records when it was last changed, along with a running weighted sum ($v_{tot}$).
Upon encountering a mislabeled instance, $v$ is updated (as required by the ``vanilla'' perceptron), $v_{tot}$ updates to include the weighted sums before the components of $v$ are changed, and the time-stamp vector is set to the current counter for all vector components that fired.
Finally, to obtain the averaged vector,  $v_{tot}$ is divided by the number of examples encountered and is returned.  
Hence, the algorithm requires minimal storage, and runs  efficiently provided the decoding, that is, the labeling algorithm, is quick.  
In our case, we employed a simple greedy model, which labels each word inductively.

As an intuitive but informal justification for the averaged perceptron, consider a scenario where 
the perceptron vector is initialized and succeeds on labeling the first 9,999 of 10,000 training examples correctly, but then mis-labels the last example and therefore changes the weight vector.  
Unfortunately, a vector that has achieved at least 99.99\% accuracy has been deselected! 
The averaged perceptron is designed to prevent overfitting and in particular to counter-act the perceptron's seeming over-weighting of the final training examples\footnote{This intuitive explanation is attributed to Hal Daum\'{e} III, \url{http://ciml.info/dl/v0_8/ciml-v0_8-ch03.pdf}.}.
While formal justification, such as convergence theorems, and theorems bounding the expectation of success on test data exist for the ``vanilla'' perceptron and voted perceptron \cite{Collins:2002:DTM:1118693.1118694, freund1999large}, to our knowledge, and as noted here \cite{goldberglearning}, no formal results have been proven for the averaged perceptron.

\subsection{Feature Selection}
Recall that our goal is to use `IOB'-tagging to collectively identify multi-word phrases, in addition to  applying the appropriate domain labels; for example, a correct labeling of an instance of  ``Internet Explorer'' is ``B: Software Product'' for ``Internet'' and ``I: Software Product'' for ``Explorer''.
Hence we view this as an iterative labeling process, first applying `IOB' labels, and secondly applying the domain labels; consequently, we train two averaged perceptron classifiers.

To develop robust features, regular expressions are used to identify words that begin with a digit, contain an interior digit, begin with a capital letter, are camel-case, are snake-case, or contain punctuation, and part-of-speech tags are applied to each word using NLTK and used as features for tagging.
Similarly, once the `IOB'-labels have been applied, they are used as features for the domain specific labeling.  
We then generate binary features as follows: 
\pagebreak
\begin{center}
\textbf{Features for `IOB'-tagging}
\end{center}
\begin{itemize}
\item Unigram features for 
	\begin{itemize}
	\item previous two, current, and following two words
	\item previous two, current and following one part of speech tags
	\item previous two `IOB'-tags
	\end{itemize}
\item Bigram features for 
	\begin{itemize}
	\item previous two 'IOB'-tags
	\item previous 'IOB'-tag \& current word
	\item previous part of speech tag \& current word
	\end{itemize}
\item Regular expressions as listed above for 
	\begin{itemize}
	\item previous two, current, and following two words
	\end{itemize}	 
\end{itemize}
Following observations made in \cite{Lal:2013}, we include gazetteer features for the labels ``Software Vendor'' and ``Software Product''; that is, sets of Software Vendor and Software Products are collected during training.  
Upon an occurrence of such a word, the appropriate gazetteer feature fires.  
\begin{center}
\textbf{Features for domain-tagging} 
\end{center}
\begin{itemize}
\item Unigram features for 
	\begin{itemize}
	\item previous two, current, and following two words
	\item previous two, current and following one part of speech tags
	\item previous two, current, and following `IOB'-tags 
	\item previous two domain labels 
	\end{itemize}
\item Bigram features for 
	\begin{itemize}
	\item previous two domain tags
	\item previous domain tag \& current word	
	\item previous 'IOB'-tag \& current word
	\item previous part of speech tag \& current word
	\end{itemize}
\item Regular expressions as listed above for
	\begin{itemize}
	\item previous two, current, and following two words
	\end{itemize}	 
\item Gazetteer features for 
	\begin{itemize}
	\item Software Product
	\item Software Vendor
	\end{itemize} 
\end{itemize}

\section{Results}
\label{sec:results}
In order to examine the performance of the taggers, five-fold random sub-sampling validation is performed on the automatically labeled corpus of NVD text, which is comprised of 15,192 text descriptions averaging about 50 words each.    
For various sizes of data samples ($n$), five random samples of $n$ text descriptions are split 80/20 \% into training and testing sets.  
For experimentation with both feature and model selection, a prototype was coded in Python, and subsequently, a faster implementation, which relied on the Apache OpenNLP library\footnote{\url{https://opennlp.apache.org}}, was developed. 
We provide both the Python code and the OpenNLP configuration details online for those interested,\footnote{\url{https://github.com/stucco/auto-labeled-corpus}} and report the performance results of the OpenNLP runs in Tables ~\ref{results_1}~,\ref{results_2}.
In particular, precision, recall, accuracy, F1-score, and training time, that is, actual clock time in seconds as observed on a  Macbook Pro with 2.3Ghz Intel quad-core i7, 8GB memory, 256GB flash storage.
We note that treating the `IOB'-tagging and the domain labeling separately allowed unambiguous analysis of the performance; that is, trying to judge accuracy of both labels at once results in cases where, for example, the `IOB'-tag is correct but the domain specific labels are incorrect, and no principled treatment exists.

Both the Python and OpenNLP implementations performed with almost perfect accuracy, with slightly better performance by the OpenNLP implementation on the domain-specific labeling, although, as expected, the OpenNLP implementation is much faster.
Perhaps the most satisfying observation in light of the result is that as the data size increases, training time seems to be growing only linearly, and as expected, precision, recall, and accuracy are monotone increasing.
Hence, in the abundance of training data, as furnished by our auto-labeling technique, state-of-the-art entity extractors can perform exceptionally in both accuracy and speed.  


The Joshi et. al. work, \cite{joshiextracting}, which trained the Stanford NER (using a CRF, a global model) for extracting very similar entities reported much more modest results, namely,  precision = .837, recall = .764, for an F1 score = .799.  
(Accuracy and training time were not recorded.)   
Moreover, we recall that Joshi et. al. used a hand-labeled training corpus of 240 CVE descriptions, 80 Microsoft or Adobe security bulletins, and 30 security blogs, a corpus of approximately one thirtieth of our full NVD data set.
As CRFs have also established themselves in the literature as state-of-the-art entity extractors, we conjecture that there are two reasons for the relatively lower performance in the Joshi paper, namely that their training set is substantially smaller than ours, and also more varied in the types of text it included.  
\begin{center}
\begin{table}[tbp]
\centering
\caption{\textbf{OpenNLP `IOB'-Labels} \label{results_1}}
\begin{threeparttable}
\begin{tabular}{cccccc}
\toprule
$n$ & P & R & F1 & A & T (sec)  \\
\midrule
500 & 0.906 & 0.929 & 0.917 & 0.944 & 1.192 \\ 
1000 & 0.921 & 0.935 & 0.928 & 0.950 & 1.396  \\ 
2500 & 0.926 & 0.966 & 0.944 & 0.962 & 3.023 \\ 
5000 & 0.947 & 0.950 & 0.948 & 0.965 & 5.468  \\ 
15192 & 0.963 & 0.968 & 0.965 & 0.976 & 15.265 \\ \bottomrule
\end{tabular}
\end{threeparttable}
\begin{tablenotes}
	\small
	\item Note: In both Tables \ref{results_1} and \ref{results_2}, $n$ refers to the number of NVD descriptions, which contain about 50 words on average.  For each $n$, five random samples are divided 80/20\% into training and test sets.  Precision, recall, F1-score, accuracy, and training time are reported.  
  \end{tablenotes}
\end{table}
\end{center}
\begin{center}
\begin{table}[tbp]
\centering
\caption{\textbf{OpenNLP Domain Labels} \label{results_2}}
\begin{threeparttable}
\begin{tabular}{cccccc}
\toprule
$n$ & P & R & F1 & A & T (sec)  \\
\midrule
100   & 0.938 & 0.918 & 0.928 & 0.952 & 0.361 \\ 
500   & 0.965 & 0.965 & 0.965 & 0.976 & 0.890 \\ 
1000 & 0.972 & 0.979 & 0.975 & 0.983 & 1.996 \\ 
2500 & 0.980 & 0.986 & 0.983 & 0.989 & 4.792 \\ 
5000 & 0.981 & 0.988 & 0.984 & 0.989 & 9.530 \\ 
15192 & 0.989 & 0.993 & 0.991 & 0.994 & 28.527 \\ \bottomrule
\end{tabular}
\end{threeparttable}
\end{table}
\end{center}

\section{Conclusion}
Our auto-labeling technique gives an expedient way to annotate unstructured text using associated structured database fields as a step towards deploying the machine learning capabilities for entity extraction to diverse and tailored applications.  
With respect to automating extraction of security specific concepts,  we provide a publicly available corpus labeled with security entities and a trained MEM for identification and classification of appropriate entites, which exhibited extremely accurate results.
Additionally,  since many sources for our auto-labeling (NVD, CVE, ...) provide RSS feeds, we seek to automate the process of acquiring and auto-labeling the new data to provide an ever growing corpus, which hopefully will help extraction methods adapt to changing language trends.
As the overall telos of this work is to accurately label ``real world'' documents containing timely security information, future work will include making the technique operationally effective by testing and tweaking the method on desired input texts.  
Lastly, upon sufficient progress towards an entity extraction system, we plan to incorporate the extraction technique into a larger architecture for acquiring documents from the web and populating a database with the domain specific concepts as an aid to security analysts.  

\section{Acknowledgments}
This material is based on research sponsored by the following: the Department of Homeland Security (DHS) Science and Technology Directorate, Cyber Security Division (DHS S\&T/CSD) via BAA 11-02; the Department of National Defence of Canada, Defence Research and Development Canada (DRDC); the Kingdom of the Netherlands; and the Department of Energy (DOE).
The views and conclusions contained herein are those of the authors and should not be interpreted as necessarily representing the official policies or endorsements, either expressed or implied, of the following:  the Department of Homeland Security; the Department of Energy; the U.S. Government; the Department of National Defence of Canada, Defence Research and Development Canada (DRDC); or the Kingdom of the Netherlands.


\bibliographystyle{IEEEtran}
\bibliography{ase-paper-references}

\end{document}